\documentclass[runningheads]{llncs}
\usepackage{amsmath}
\usepackage{tabto}
\usepackage{graphicx,url}
\usepackage[table,xcdraw]{xcolor}
\usepackage [english]{babel}
\usepackage [autostyle, english = american]{csquotes}

\def \pranav [#1]{\textcolor{red}{p: #1}}
\def \nagaraj [#1]{\textcolor{blue}{N: #1}}
\begin{document}
%
% \title{Contribution Title\thanks{Supported by organization x.}}
\title{Context-based out-of-vocabulary word recovery for ASR systems in Indian languages}
\titlerunning{Context-based OOV word recovery}
\author{Arun Baby, Saranya Vinnaitherthan, Akhil Kerhalkar, Pranav Jawale,\\ Sharath Adavanne, Nagaraj Adiga}
% %
\authorrunning{Arun Baby et al.}

\institute{Zapr Media Labs, India}

%  \\
% \email{arunbaby0@gmail.com}

\maketitle              % typeset the header of the contribution
\begin{abstract}
Detecting and recovering out-of-vocabulary (OOV) words is always challenging for Automatic Speech Recognition (ASR) systems. Many existing methods focus on modeling OOV words by modifying acoustic and language models and integrating context words cleverly into models. To train such complex models, we need a large amount of data with context words, additional training time, and increased model size. However, after getting the ASR transcription to recover context-based OOV words, the post-processing method has not been explored much. In this work, we propose a post-processing technique to improve the performance of context-based OOV recovery. We created an acoustically boosted language model with a sub-graph made at phone level with an OOV words list. We proposed two methods to determine a suitable cost function to retrieve the OOV words based on the context. The cost function is defined based on phonetic and acoustic knowledge for matching and recovering the correct context words in the decode. The effectiveness of the proposed cost function is evaluated at both word-level and sentence-level. The evaluation results show that this approach can recover an average of 50\% context-based OOV words across multiple categories.

\end{abstract}

\section{Introduction}

%Word error rate has been used as the primary measure of performance for any ASR models. The assumption here is that all the words are equally important. However, in an utterance there are important keywords that dictates the entire meaning or usefulness of the sentence, especially in case of conversational AI. This makes WER a poor measure for sentences with some special context. 
Detecting out-of-vocabulary (OOV) words, especially proper nouns, has always been challenging for any Automatic Speech Recognition (ASR) system. Capturing these nuances in the language model is challenging as these important context words are mostly unavailable in the training data and result in a high word error rate (WER). Moreover, the downstream tasks that depend on these keywords are also affected significantly because of the wrong recognition. 

Consider the following conversation between two people.

% \pranav[next long sentence, can we say, Low or zero frequency of such words in training data, increases WER, as well as the downstream tasks heavily dependent on identification on these keywords get adversely affected.]{}
\quad Person 1: ``Where do you live?" \

\quad Person 2: ``I live in Brno"

Now assume that we are using two different ASR models to decode the utterance of Person 2.

\quad Model 1: ``I live in beer" \

\quad Model 2: ``* * in Brno"

Using WER as the primary metric for evaluating these models, Model 1 gives 25\% WER, whereas Model 2 gives 50\%. The low WER will imply that Model 1 is better than Model 2. However, in real-life use cases, the recognition result of Model 2 is far more helpful. Lower WER for specifically important keywords is better than overall WER. In conversational AI, downstream tasks like intent/entity recognition solely depend on detecting important keywords or OOV words. When we say OOV, we refer to a list of words and phrases for which we don't have sentences containing these words available in the language model (LM) training data. In traditional OOV recovery tasks, the OOV words are unknown at training and decoding. However, in the case of conversational AI, we can have a list of expected words while decoding because of the known context of the conversation. However, training a general LM with all the OOV words is not feasible due to difficulty in obtaining the sentences with OOV words. So we are trying to recover known OOV words while decoding but not part of the normal LM.

Getting text and audio pairs is difficult in the case of the OOV words. The task of creating meaningful sentences with OOV words is laborious. There are multiple methods to model the OOV context words in the literature. One solution is implicit acoustic model training using synthetic data consisting of OOV words~\cite{9414778}. Another solution is dynamic graph manipulation, which includes recompiling graphs on the fly. However, graph creation is a resource-hungry task. One of the existing approaches uses a hybrid language model with word and subword-level language models. Then a phoneme-to-grapheme model is used to recover OOV words~\cite{kombrink2010recovery}. The OOV recovery heavily depends on the subword-level graph in this approach. This method performs a basic OOV recovery and does not use context information. A phone confusion-based matching approach for retrieving song names with context is explored in ~\cite{chaudhari2007improvements}. This model use an index of n-grams from the documents to find the phonetic matching with the target search. Here, the search is not context-based and uses a vanilla edit distance for matching.  In~\cite{shah2020cross} a common phoneme label set-based approach is proposed but does not focus on improving the cost function to compute the better phone similarity.
\cite{chineese9362062} uses character-based modeling and graph manipulations to recover OOV words. However, while recovering the OOV words, context words are not used. A clever training technique is used to improve the OOV modeling in ~\cite{kim2021improving}, which makes use of the existing corpus itself, which enhances the low-frequency words recovery in the training data. A multiple-level, namely, word, phonetic, and grapheme-level matching-based recovery of name entity recognition is proposed in ~\cite{garg2020hierarchical} but uses vanilla distance measure for computing phone similarity.

Most of the approaches above are developed for English languages or monolingual systems. In all the above methods, the focus is on improving the modeling technique or post-processing with multiple blocks. The phone confusion matrices used in phonetic matching use vanilla distance measures. Our approach explores various phoneme level cost measures to post-process the decoded text and match it with a context word list. We have found that using fixed cost for OOV recovery deteriorates the performance as acoustically similar phoneme confusions are penalized heavily. We have developed strategies to formulate a suitable phone confusion matrix for OOV recovery. We use two different ASR LM models. We get the standard hypothesis with the first LM and the second LM acoustically boosted to obtain the OOV words. The output of the second model is used in recovering the OOV words with the help of phone sequences. We merge the hypothesis to have a better transcription based on the word level timing information of both the decoded output. We evaluated the proposed method with the word- and sentence-level data sets. The word-level data set consists of synthesized words from a bilingual Text-to-Speech (TTS) system. The sentence-level data set consists of telephonic conversational data collected internally in our organization. We showed that the proposed method performs better context-based OOV recovery at both word and sentence-level over baseline method.

%Conversational AI can use context words list to have better recognition, which will improve the system's overall usability. This paper has attempted to improve these systems to have better recognition than conventional systems. We focus mainly on post-processing, where the decoding is enhanced with the help of better phoneme confusion matrices.

\section{Proposed Method}
\begin{figure}[h]
    \centering
    \centerline{\includegraphics[width=\columnwidth, keepaspectratio]{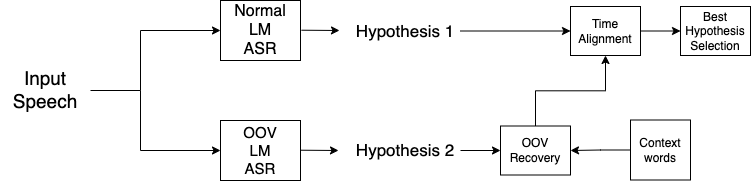}}
    \caption{Block diagram of the proposed context-based OOV word recovery approach.}
    \label{img:basic}
\end{figure}

\begin{figure}[!h]
    \centering
    \centerline{\includegraphics[width=\columnwidth, keepaspectratio]{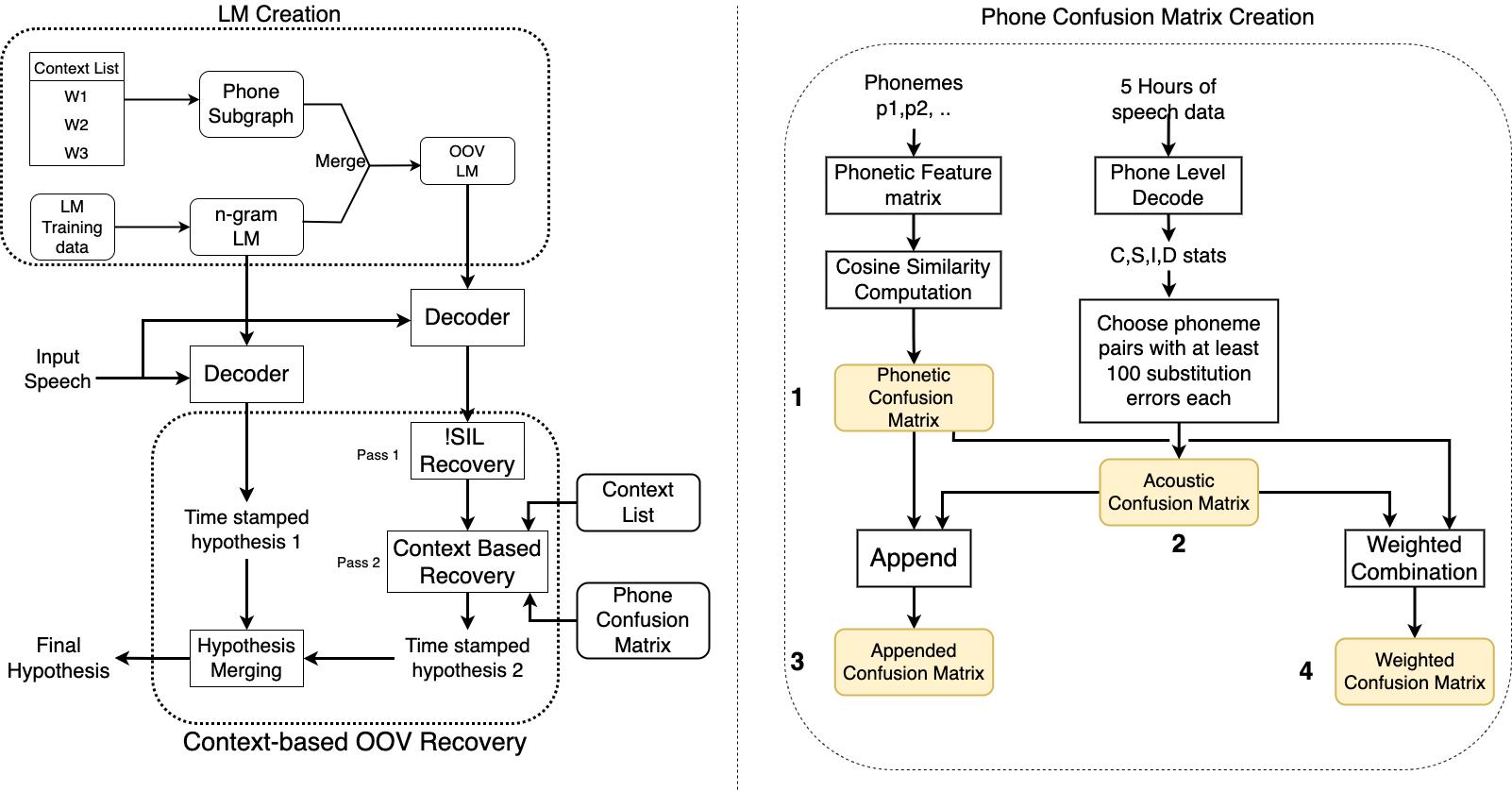}}
    \caption{The detailed diagram of the proposed approach consists of three main blocks: LM creation, Phone confusion matrix creation, and Context-based OOV recovery blocks respectively. We used four different confusion matrix creation methods: 1. Phonetic, 2. Acoustic, 3. Append, and 4. Weighted-based methods are highlighted with an yellow color box.}
    \label{img:detailed}
\end{figure}

In the proposed approach, the context-based OOV recovery is made with the help of multiple modules. Figure~\ref{img:basic} shows the basic flow of OOV recovery using the proposed approach. Here we use two different ASR models for decoding the same speech signal. The one ASR model does conventional decoding using normal LM. The second ASR model accomplishes context words detection using OOV LM. Then decoded text from both the normal and OOV LMs are time-aligned. The OOV words are recovered based on the phone similarity between the context word list and decoded hypothesis from ASR. The main focus of our approach is obtaining a suitable phone cost function. Figure~\ref{img:detailed} shows the detailed flow of the proposed method. There are three different blocks. The first is the LM creation block, where specific graph level manipulations are performed for possible OOV recovery. The second block is for defining the cost function based on the phone confusion matrix, and the final block is the OOV recovery block. Each block is explained in detail further below.

\subsection{LM creation}

Our proposed method uses two ASR systems to decode the same data and combine the hypothesis based on the recovered OOV words. The acoustic model (AM) remains the same in both models, whereas the LMs are different. The first is the normal LM, and the second is an OOV-LM. Here n-gram~\cite{enwiki:1073019765} based LM, a probabilistic model created with an entire training corpus. The normal LM consists of a graph for phonemes, silence (SIL), and unknown words (!SIL). The normal LM does not contain any sentences with OOV words. %The OOV-LM is created with the context words list to make a phone level finite-state automaton graph.
The !SIL symbol is used to model unknown words in case of normal LM. In the OOV LM, !SIL symbol is replaced with the phone level sub-graph. A 4-gram phone model is created using all the context words to form a phone-level sub-graph. This graph is connected to all other nodes. It ensures that any OOV can be recognized before/after any other word. The graph scale controls the path probability of the sub-graph. To recognize OOV words, graph scale of !SIL node is increased manyfold. Additionally, while decoding with the OOV-LM, the acoustic scale of the graph is increased to emphasize the phoneme level output.

\begin{figure}[!h]s
    \centering
    \centerline{\includegraphics[width=\columnwidth, keepaspectratio]{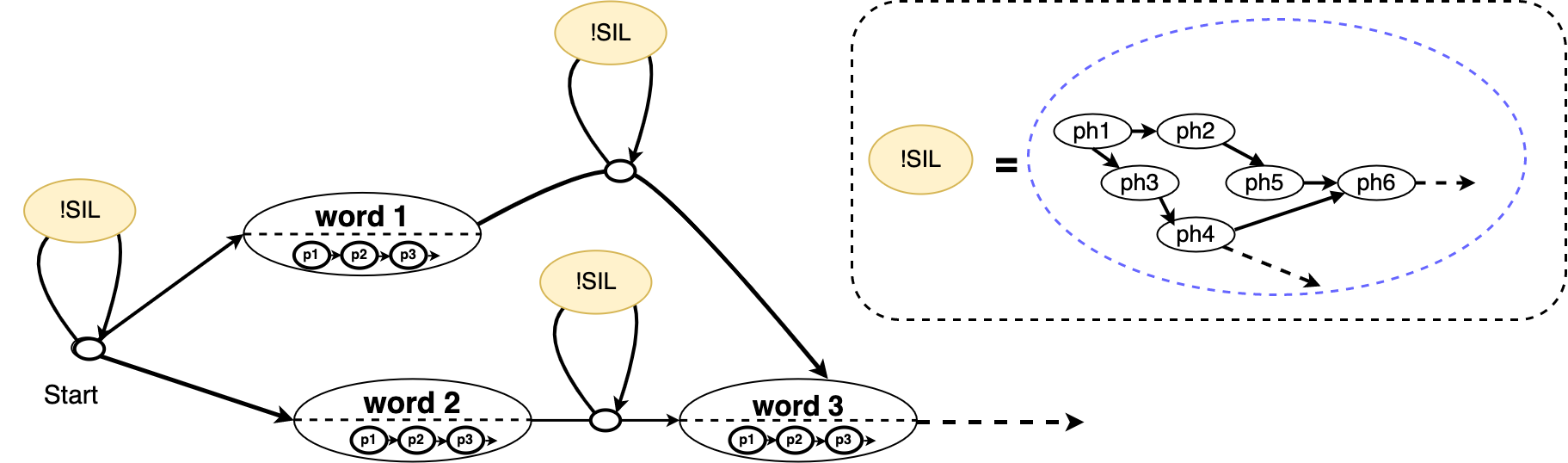}}
    \caption{The sample view LM graph consists of multiple words and !SIL nodes on the left side image, and its detailed view of phone sub-graph shown for one !SIL node on right side image, respectively.}
    \label{img:oovlm}
\end{figure}

A sample OOV-LM is shown in Figure~\ref{img:oovlm}. An n-gram word LM is depicted with !SIL node attached to each word boundary. The sub-graph structure of the !SIL is shown on the right side of the figure. Here we can see that !SIL node is a complex phoneme level sub-graph with interconnections based on the OOV context word list.

\subsection{Phone confusion matrix creation}
\label{ssec:confusion} 
The main contribution of this work is the cost function formulation-based on phone confusion matrix. The cost function is critical for good performance for any match and replacing algorithm. The distance metric used here decides the overall quality of the approach. As the recognized phoneme sequence for OOV words is likely to have some errors, we need to find the best matching OOV word using a distance metric. We compare the ideal phone sequence of context words/phrases with the decoded phone sequence. The Levenshtein  distance~\cite{enwiki:1067984406} between the two phone sequences is used to match the word similarity. The vanilla Levenshtein distance metric gives equal weightage between phone insertions, deletions, and substitutions. It reduces the system performance because the distance between two acoustically similar phonemes is the same as between different phonemes. We explored various techniques to arrive at a better phoneme distance metric for tackling this.

The Levenshtein distance algorithm is defined as:
\begin{equation}
\label{eq:ldist}
c_{i,j} = min\left\{\begin{matrix}
c_{i-1,j}+D_{cost} \\
c_{i,j-1}+I_{cost} \\ 
c_{i-1,j-1}+S_{cost} \\
\end{matrix}\right.
\end{equation}
where, $c_{i,j}$ is the cost at $i^{th},j^{th}$ location of search matrix, is calculated recursively according to the Equation~\ref{eq:ldist}, $D_{cost}$ is the deletion cost,  $I_{cost}$ is the insertion cost and $S_{cost}$ is the substitution cost. The Equation~\ref{eq:ldist} computes the cost distance between two phoneme sequences. For the vanilla Levenshtein algorithm, $D_{cost}$ = $S_{cost}$ = $I_{cost}$ = 1. We are experimenting with the substitution cost in this work, which can be considered a phone confusion cost. The insertion and deletion costs are kept constant, whereas the substitution cost is adjusted based on different methods for optimal performance. The vanilla algorithm also keeps the substitution cost as a constant value (denoted as Hard cost) for the baseline system. In this work, we explored phonetic and acoustic similarity-based phone cost function.

% \pranav[do we need more explanation of L dist algorithm? or is it assumed that readers already know it? For example, we should say that c i,j is the cost at i,jth search matrix place. We are not defining i and j in prev. paragraph..  A: L dist should be known. Also we are giving ref to the L dist.]{}

\subsubsection{1. Phonetic similarity:}
The phonetic similarity cost function is based on the property of the phones. The nearest set of phones will have a lesser cost, and this can be considered a substitution cost, as it is a distance measured between the set of phonemes. Here we figure out the distance between two phones based on the phonetic similarity. A total of 39 different properties for each phoneme are identified. Some of these include properties of vowel (short, medium, and long), consonant (plosive, fricative, Central-approximant,  Lateral-approximant, flap, velar,  palatal, retroflex, dental, labial, aspirated, etc.), nukta, halanta, and anusvara. For each phoneme, we create a vector based on the above phonetic attributes, and more details about phonetic similarity are mentioned in ~\cite{kunchukuttan2020indicnlp}. The confusion cost between two phones is the cosine distance between the vector representation. It is shown as $1$ in the confusion matrix creation block of Figure~\ref{img:detailed}. A total of 4225 confusion combinations are obtained here for 65 unique phones in the phone set we used~\cite{baby21_ssw}. Basically, with phonetic similarity cost function we can get cost function for all phone pairs.

%These  are:  vowel, consonant, nukta, halanta, anusvara, misc \pranav[what's misc? A:Don't know actually. should we only list a few rather that all 39?]{}, short vowel, long vowel, weak, medium, strong, independent vowel, dependent vowel, plosive, fricative, Central-approximant,  Lateral-approximant, flap, velar,  palatal, retroflex, dental, labial, aspirated, not aspirated, voiced, unvoiced, nasal, not nasal, front, central, back, close,  close-mid, open-mid, open, rounded, not rounded. \pranav[aren't some of these attributes redundant? when we have ``nasal" attribute, why do we need a separate ``non nasal" att.?  A: should we skip mentioning each of these, since a reference is given? ]{}

\subsubsection{2. Acoustic similarity:}
The acoustic similarity between two phonemes is the similarity in the auditory level. Since the phonetic similarity might not capture the actual acoustic likeliness in the training data, we tried to develop a metric based on the ASR decoded data. A corpus of 5 hours of bilingual (Hindi-English) speech is decoded using the phone level language model to determine the acoustic similarities. Code-mixing across languages is predominant in the case of Indian languages~\cite{Thomas2018}. So to obtain good phone confusion across both languages, we used bilingual ASR.  Next, we get the substitution, deletion, insertion, and correct hypothesis statistics for each reference phoneme in the test data. For calculating the acoustic confusion, all the correct hypotheses and the substitutions are considered. Acoustic similarity cost is defined as:

% \begin{equation}
%  confusion = (1 - (C/(C+S)))^4
% \end{equation}

%\begin{equation}
% acoustic\,similarity\,cost = (1 - (N_C/(N_C+N_S)))^4
%\end{equation}

\begin{equation}
(1 - (N_C/(N_C+N_S)))^4
\end{equation}
where, $N_C$ is the number of phonemes correctly recognized and $N_S$ the number of phonemes substituted in the decoded data. Here, this ratio is raised to the power of $4$ to have a broader range of values between $0$ and $1$. Only phones with more than $100$ substitutions are considered in our experiment. After applying these conditions, around $136$ unique frequently occurring phone confusion mappings are found. This process is shown as $2$ in the confusion matrix creation block of Figure~\ref{img:detailed}. 

\subsubsection{3 \& 4. Ensemble cost function:} 
To obtain better systems, we have tried to combine the above two cost computation methods in two different ways. In the first ensemble method  ("Append"), phone confusion pairs that are not part of the acoustic similarity matrix are added from the phonetic similarity. This method is shown as $3$ in the confusion matrix creation block of Figure~\ref{img:detailed}. In the second method ("Weighted"), both acoustic and phonetic similarity methods are combined in a weighted manner by giving equal weighting to both the functions, and this process is shown as $4$ in the confusion matrix creation block of Figure~\ref{img:detailed}.

\subsection{Context-based OOV recovery}
\label{ssec:OOV_recovery}
In this block, a two-pass OOV recovery is employed. The hypothesis from each of the decoders can have !SIL in the decode or some other acoustically similar words decoded. These are recovered with the help of phone confusion matrices. The hypothesis from the ASR with OOV LM model is searched for the !SIL token, which is the phone-only path in the graph. The corresponding phone sequence is recovered from the graph if this token exists. In the first phase, this sequence is checked against the list of context words. If there is a direct phone matching, the context word is recovered, and this recovery process is denoted as Pass $1$. In the second phase, a phone-based matching is performed across the decoded sentence. In this case, the variable window size is chosen to match the parts of the decoded sentence into any of the context word lists. Each decoded word is compared phonetically with the context words to find the appropriate matching within a given window time interval. The one with the least confusion cost is replaced from the context list. However, a phone confusion threshold cost restricts this replacement if the confusion is too high. This process recovery is denoted as Pass $2$. Then both the OOV recovered sentence, and the normal LM decoded text is time-aligned. The context words, if recovered, are merged back into the actual decode with the help of this time alignment. Sometimes overlap can happen, and we keep both the words in those cases.

\section{Experimental setup}
The ASR experiments were conducted on synthetic and telephonic conversational data collected internally for research purposes. 
We trained the ASR model using this data at an 8 kHz sampling rate.
Three different acoustic models (English, Hindi, and Hinglish) are trained in tandem with the LM to analyze if any language dependency is there for recovering OOV context words. The  English, Hindi, and Hinglish AM models are trained using 500, 342, and 842 hours of data, respectively. The normal LM model is trained with 80 k conversational text sentences. For OOV LM, along with sentences used for normal LM, a sub-graph is created with a context word list and more details mentioned in section~\ref{ssec:OOV_dataset}.
We use Kaldi toolkit~\cite{povey2011kaldi} for training AM as well as decoding and SRILM toolkit~\cite{stolcke2002srilm} for training LM. The unified-parser~\cite{nlp:tsd16conf} is used as the phonemizer for all the experiments. A common phone set across Hindi and English is used further for obtaining a language-agnostic phoneset~\cite{baby21_ssw}.

After OOV LM is created, the post-processing method using different phone confusion matrices is tested out as explained in section~\ref{ssec:confusion}. Here the hard cost is compared against other cost functions. We are trying to find matching words with roughly the same number of phonemes and mainly differing in similar-sounding phonemes substituted for each other. We found that the heuristic constraint mentioned in Equation~\ref{eq:s_cost} works well while fixing the insertion and deletion costs for all the cases. The logic behind being the substitution cost ($S_{cost}$) should be lesser than the sum of insertion cost ($I_{cost}$) and deletion cost ($D_{cost}$) for the Levenshtein distance algorithm to choose the substitution path.  %\nagaraj[you can mention based on our experiments or you can cite paper if you are taking it from other source A: given previous sentence for justifications, there is not paper to cite afaik]{}
\begin{equation}
\label{eq:s_cost}
  S_{cost} < I_{cost} + D_{cost}  
\end{equation}

In the case of sentence-level testing, the objective is twofold. The first is a recovery of the OOV word. The second is to identify if the detected OOV word is in the expected position to reference transcription. Force alignment is performed on the test data to get ground truth time stamps for each word in transcription. While evaluating, the slight margin of error is allowed using a time windowing approach on the word boundary's start and end points. Different window sizes are also analyzed to find the optimal hyper-parameter. On top of this, different merge cost thresholds are also tested out. The error allowance (total phone confusion cost) is adjusted based on the maximum threshold of merging cost.

\subsection{Data set for context-based OOV recovery}
\label{ssec:OOV_dataset}
Experiments are performed on two different data sets. The first one is a synthesized speech word/phrase level data set. The second one consists of sentence utterances extracted from a natural conversation. The first data set (D1) consists of OOVs words in three different categories like state, city, and car dealership names collected from a conversational voice bot. The size of the OOV context words for the state, city, and dealer names is 27, 370, and 546, respectively. The state and city names are mostly Indian-origin proper nouns written in English, and the dealer names are a good mixture of Hindi and English combinations. A subset of these context words is shown in Table~\ref{tab:OOV_list_sample}, as part of the D1 column. These entities are synthesized using a custom bilingual (Hindi-English) TTS system\cite{baby21_ssw}. The TTS model is built using Espnet2~\cite{hayashi2019espnettts} and parallelWaveGaN~\cite{yamamoto2020parallel} tools. The Espnet toolkit is used as a front end to convert text to spectrogram with Tacotron 2 (v3) recipe~\cite{hayashi2019espnettts}. The Mel spectrogram output of front end mapped to waveform using the parallel wavegan (PWG) vocoder. The TTS system is trained with Indic TTS~\cite{babycbblr2016} as the data set. Both Hindi and English languages consist of 10 hours from the same speaker to create this TTS model.

\begin{table}[]
\caption{A subset of the OOV context word list used in our experiments: The word-level (D1) data set consists of three categories, namely, state, city, and car dealership names respectively; the sentence-level (D2) data set consists of conversational data.}
\label{tab:OOV_list_sample}
\centering
\resizebox{\textwidth}{!}{%
\begin{tabular}{|ccc|c|}
\hline
\multicolumn{3}{|c|}{\textbf{Word-level (D1) }}                                                                     & \textbf{Sentence-level (D2)} \\ \hline
\multicolumn{1}{|c|}{\textbf{States}} & \multicolumn{1}{c|}{\textbf{Cites}} & \textbf{Dealer names}                                           & \textbf{Conversational} \\ \hline
\multicolumn{1}{|c|}{Andhra pradesh} & \multicolumn{1}{c|}{Agartala}      & \cellcolor[HTML]{FFFFFF}Aarav automobiles company         & Maruthi                 \\ \hline
\multicolumn{1}{|c|}{Bengal}         & \multicolumn{1}{c|}{Baripada}      & \cellcolor[HTML]{FFFFFF}Akvee automotives private limited & Manjunath               \\ \hline
\multicolumn{1}{|c|}{Chattisgarh} & \multicolumn{1}{c|}{Berhampore} & India garage bangalore          & Kamanahalli \\ \hline
\multicolumn{1}{|c|}{Gujarat}     & \multicolumn{1}{c|}{Dehradun}   & Jhalrapatan                     & Berhampur   \\ \hline
\multicolumn{1}{|c|}{Jharkhand}   & \multicolumn{1}{c|}{Kumarhatti} & Maa bindheswani automobile      & Dimple      \\ \hline
\multicolumn{1}{|c|}{Kerala}      & \multicolumn{1}{c|}{Navashahar} & Saakaars                        & Kalpesh     \\ \hline
\multicolumn{1}{|c|}{Maharashtra}    & \multicolumn{1}{c|}{Samastipur}    & Somya vehicles private limited                            & Mahindra                \\ \hline
\multicolumn{1}{|c|}{Punjab}      & \multicolumn{1}{c|}{Sultanpur}  & Trendy wheels                   & Ahmedabad   \\ \hline
\multicolumn{1}{|c|}{Telangana}   & \multicolumn{1}{c|}{Warangal}   & Vaishnavi carz                  & Corona      \\ \hline
\multicolumn{1}{|c|}{Uttarakhand} & \multicolumn{1}{c|}{Zirakpur}   & Zulaikha motors private limited & Sarita      \\ \hline
\end{tabular}%
}
\end{table}

The second data set (D2) uses internally collected telephonic conversational data. The speakers here are native Hindi bilinguals. The average duration of these sentences is around 5 seconds. Given a list of OOV words, we identify sentences that contain them. About 50 sentences are chosen based on a selection of 15 OOV words. A subset of the OOV words chosen is shown in Table~\ref{tab:OOV_list_sample} as a D2 column. 

\section{Results and Analysis}
Two experiments are conducted on data set D1. In the first case, the entire OOV words as a context list are passed to the OOV recovery module to recover the correct one (Experiment $1$). Only the right OOV word is given to the OOV recovery module (Experiment $2$) in the second case. The results of the word-level experiments denoted as D1 are shown in Table~\ref{tab:word_results}. Five different cost functions are tested here as explained in Section~\ref{ssec:confusion}. Pass $1$ and Pass $2$ denote the number of words recovered in each pass with the corresponding modules described in Section~\ref{ssec:OOV_recovery}. Three different acoustic models developed with English, Hindi, and Hinglish data are tested by keeping the LM same. The table shows that the Hinglish ASR model performed better than the other two acoustic models. It may be because more hours of data are used while training the Hinglish ASR than in the other two models. Here we could see that the cost function based on Acoustic similarity performed better than phonetic similarity in cases where the OOV word list is more. The best OOV recovery of individual categories of states, cities, and dealer names is 85\%,  59\%, and 39\%, respectively, highlighted as bold font in Table~\ref{tab:word_results}. Another observation is that as the number of words in context list increases (dealer category), the system's performance decreases.

% \caption{Experiment 1: OOV recovery rate of multiple cost functions for the D1 data set with a multiple OOV context word list}
% \label{tab:word_results}
% \centering

\begin{table}[]
\caption{Experiment 1: OOV recovery rate of multiple cost functions for the D1 data set with a multiple OOV context word list}
\label{tab:word_results}
\centering
\begin{tabular}{|
>{\columncolor[HTML]{4A86E8}}c cccccc|}
\hline
\multicolumn{1}{|c|}{\cellcolor[HTML]{FBBC04}AM models} &
  \multicolumn{2}{c|}{\cellcolor[HTML]{FBBC04}English ASR} &
  \multicolumn{2}{c|}{\cellcolor[HTML]{FBBC04}Hindi ASR} &
  \multicolumn{2}{c|}{\cellcolor[HTML]{FBBC04}Hinglish ASR} \\ \hline
\multicolumn{1}{|l|}{\cellcolor[HTML]{4285F4}Cost Function} &
  \multicolumn{1}{c|}{Pass 1 (\%)} &
  \multicolumn{1}{c|}{Pass 2 (\%)} &
  \multicolumn{1}{c|}{Pass 1 (\%)} &
  \multicolumn{1}{c|}{Pass 2 (\%)} &
  \multicolumn{1}{c|}{Pass 1 (\%)} &
  Pass 2 (\%) \\ \hline
\multicolumn{7}{|c|}{\cellcolor[HTML]{4A86E8}\textbf{State Category (27 OOV words)}} \\ \hline
\multicolumn{1}{|c|}{\cellcolor[HTML]{4A86E8}{\color[HTML]{091E42} Hard cost}} &
  {\color[HTML]{091E42} 33.3} &
  \multicolumn{1}{c|}{{\color[HTML]{091E42} 40.7}} &
  {\color[HTML]{091E42} 29.6} &
  \multicolumn{1}{c|}{{\color[HTML]{091E42} 48.1}} &
  {\color[HTML]{091E42} 37} &
  {\color[HTML]{091E42} 48.1} \\ \cline{1-1}
\multicolumn{1}{|c|}{\cellcolor[HTML]{4A86E8}{\color[HTML]{091E42} Acoustic}} &
  {\color[HTML]{091E42} 33.3} &
  \multicolumn{1}{c|}{{\color[HTML]{091E42} 44.4}} &
  {\color[HTML]{091E42} 29.6} &
  \multicolumn{1}{c|}{{\color[HTML]{091E42} 59.3}} &
  {\color[HTML]{091E42} \textbf{37}} &
  {\color[HTML]{091E42} \textbf{48.1}} \\ \cline{1-1}
\multicolumn{1}{|c|}{\cellcolor[HTML]{4A86E8}{\color[HTML]{091E42} Phonetic}} &
  {\color[HTML]{091E42} 33.3} &
  \multicolumn{1}{c|}{{\color[HTML]{091E42} 51.9}} &
  {\color[HTML]{091E42} 29.6} &
  \multicolumn{1}{c|}{{\color[HTML]{091E42} 48.1}} &
  {\color[HTML]{091E42} 37} &
  {\color[HTML]{091E42} 37} \\ \cline{1-1}
\multicolumn{1}{|c|}{\cellcolor[HTML]{4A86E8}{\color[HTML]{091E42} Append}} &
  {\color[HTML]{091E42} 33.3} &
  \multicolumn{1}{c|}{{\color[HTML]{091E42} 48.1}} &
  {\color[HTML]{091E42} 29.6} &
  \multicolumn{1}{c|}{{\color[HTML]{091E42} 48.1}} &
  {\color[HTML]{091E42} 37} &
  {\color[HTML]{091E42} 37} \\ \cline{1-1}
\multicolumn{1}{|c|}{\cellcolor[HTML]{4A86E8}{\color[HTML]{091E42} Weighted}} &
  {\color[HTML]{091E42} 33.3} &
  \multicolumn{1}{c|}{{\color[HTML]{091E42} 51.9}} &
  {\color[HTML]{091E42} 29.6} &
  \multicolumn{1}{c|}{{\color[HTML]{091E42} 48.1}} &
  {\color[HTML]{091E42} 37} &
  {\color[HTML]{091E42} 37} \\ \hline
\multicolumn{7}{|c|}{\cellcolor[HTML]{4A86E8}\textbf{City Category (370 OOV words)}} \\ \hline
\multicolumn{1}{|c|}{\cellcolor[HTML]{4A86E8}{\color[HTML]{091E42} Hard cost}} &
  {\color[HTML]{091E42} 6.5} &
  \multicolumn{1}{c|}{{\color[HTML]{091E42} 25.7}} &
  {\color[HTML]{091E42} 12.4} &
  \multicolumn{1}{c|}{{\color[HTML]{091E42} 27.6}} &
  {\color[HTML]{091E42} 11.6} &
  {\color[HTML]{091E42} 29.7} \\ \cline{1-1}
\multicolumn{1}{|c|}{\cellcolor[HTML]{4A86E8}{\color[HTML]{091E42} Acoustic}} &
  {\color[HTML]{091E42} 6.5} &
  \multicolumn{1}{c|}{{\color[HTML]{091E42} 47}} &
  {\color[HTML]{091E42} 12.4} &
  \multicolumn{1}{c|}{{\color[HTML]{091E42} 41.6}} &
  {\color[HTML]{091E42} \textbf{11.6}} &
  {\color[HTML]{091E42} \textbf{47}} \\ \cline{1-1}
\multicolumn{1}{|c|}{\cellcolor[HTML]{4A86E8}{\color[HTML]{091E42} Phonetic}} &
  {\color[HTML]{091E42} 6.5} &
  \multicolumn{1}{c|}{{\color[HTML]{091E42} 25.7}} &
  {\color[HTML]{091E42} 12.4} &
  \multicolumn{1}{c|}{{\color[HTML]{091E42} 24.6}} &
  {\color[HTML]{091E42} 11.6} &
  {\color[HTML]{091E42} 27} \\ \cline{1-1}
\multicolumn{1}{|c|}{\cellcolor[HTML]{4A86E8}{\color[HTML]{091E42} Append}} &
  {\color[HTML]{091E42} 6.5} &
  \multicolumn{1}{c|}{{\color[HTML]{091E42} 25.1}} &
  {\color[HTML]{091E42} 12.4} &
  \multicolumn{1}{c|}{{\color[HTML]{091E42} 24.3}} &
  {\color[HTML]{091E42} 11.6} &
  {\color[HTML]{091E42} 27.6} \\ \cline{1-1}
\multicolumn{1}{|c|}{\cellcolor[HTML]{4A86E8}{\color[HTML]{091E42} Weighted}} &
  {\color[HTML]{091E42} 6.5} &
  \multicolumn{1}{c|}{{\color[HTML]{091E42} 25.9}} &
  {\color[HTML]{091E42} 12.4} &
  \multicolumn{1}{c|}{{\color[HTML]{091E42} 24.9}} &
  {\color[HTML]{091E42} 11.6} &
  {\color[HTML]{091E42} 28.4} \\ \hline
\multicolumn{7}{|c|}{\cellcolor[HTML]{4A86E8}\textbf{Dealer Category (546 OOV words)}} \\ \hline
\multicolumn{1}{|c|}{\cellcolor[HTML]{4A86E8}{\color[HTML]{091E42} Hard cost}} &
  {\color[HTML]{091E42} 9} &
  \multicolumn{1}{c|}{{\color[HTML]{091E42} 17.8}} &
  {\color[HTML]{091E42} 0.7} &
  \multicolumn{1}{c|}{{\color[HTML]{091E42} 2.7}} &
  {\color[HTML]{091E42} 8.6} &
  {\color[HTML]{091E42} 17.2} \\ \cline{1-1}
\multicolumn{1}{|c|}{\cellcolor[HTML]{4A86E8}{\color[HTML]{091E42} Acoustic}} &
  {\color[HTML]{091E42} \textbf{9}} &
  \multicolumn{1}{c|}{{\color[HTML]{091E42} \textbf{29.9}}} &
  {\color[HTML]{091E42} 0.7} &
  \multicolumn{1}{c|}{{\color[HTML]{091E42} 7.5}} &
  {\color[HTML]{091E42} 8.6} &
  {\color[HTML]{091E42} 27.8} \\ \cline{1-1}
\multicolumn{1}{|c|}{\cellcolor[HTML]{4A86E8}{\color[HTML]{091E42} Phonetic}} &
  {\color[HTML]{091E42} 9} &
  \multicolumn{1}{c|}{{\color[HTML]{091E42} 15.6}} &
  {\color[HTML]{091E42} 0.7} &
  \multicolumn{1}{c|}{{\color[HTML]{091E42} 3.5}} &
  {\color[HTML]{091E42} 8.6} &
  {\color[HTML]{091E42} 14.5} \\ \cline{1-1}
\multicolumn{1}{|c|}{\cellcolor[HTML]{4A86E8}{\color[HTML]{091E42} Append}} &
  {\color[HTML]{091E42} 9} &
  \multicolumn{1}{c|}{{\color[HTML]{091E42} 18.3}} &
  {\color[HTML]{091E42} 0.7} &
  \multicolumn{1}{c|}{{\color[HTML]{091E42} 4.6}} &
  {\color[HTML]{091E42} 8.6} &
  {\color[HTML]{091E42} 16.5} \\ \cline{1-1}
\multicolumn{1}{|c|}{\cellcolor[HTML]{4A86E8}{\color[HTML]{091E42} Weighted}} &
  {\color[HTML]{091E42} 9} &
  \multicolumn{1}{c|}{{\color[HTML]{091E42} 17.8}} &
  {\color[HTML]{091E42} 0.7} &
  \multicolumn{1}{c|}{{\color[HTML]{091E42} 4.2}} &
  {\color[HTML]{091E42} 8.6} &
  {\color[HTML]{091E42} 16.5} \\ \hline
\end{tabular}
\end{table}

It is a common scenario in conversational call center audio to verify if the context name or location name is present in the audio or not. It can be considered a verification problem of context-based OOV recovery. In the second experiment for the D1 data set, a single correct OOV word is passed instead of the entire OOV context words list. The  OOV context words are recovered with a confidence score based on the cost criteria. The result of the same is shown in Table~\ref{tab:word_results_single}. In this case, many words are verified correctly compared to Experiment 1. Also, we could observe that cost function-based Phonetic similarity is performing better than other cost functions. It also shows that when the number of the context word list is less phonetic cost function performed better than acoustic similarity. In individual categories,  state, city, and dealer names, the best OOV verification rate is 100\%,  89\%, and 99\%, respectively, which is highlighted in bold font in the Table~\ref{tab:word_results_single}.

\begin{table}[]
\caption{Experiment 2: OOV verification results of multiple cost functions for the D1 data set with a single OOV context word}
\label{tab:word_results_single}
\centering
\begin{tabular}{|
>{\columncolor[HTML]{4A86E8}}c cccccc|}
\hline
\multicolumn{1}{|c|}{\cellcolor[HTML]{FBBC04}{\color[HTML]{091E42} AM models}} &
  \multicolumn{2}{c|}{\cellcolor[HTML]{FBBC04}{\color[HTML]{091E42} English ASR}} &
  \multicolumn{2}{c|}{\cellcolor[HTML]{FBBC04}{\color[HTML]{091E42} Hindi ASR}} &
  \multicolumn{2}{c|}{\cellcolor[HTML]{FBBC04}{\color[HTML]{091E42} Hinglish ASR}} \\ \hline
\multicolumn{1}{|l|}{\cellcolor[HTML]{4285F4}Cost Function} &
  \multicolumn{1}{c|}{Pass 1 (\%)} &
  \multicolumn{1}{c|}{Pass 2 (\%)} &
  \multicolumn{1}{c|}{Pass 1 (\%)} &
  \multicolumn{1}{c|}{Pass 2 (\%)} &
  \multicolumn{1}{c|}{Pass 1 (\%)} &
  Pass 2 (\%) \\ \hline
\multicolumn{7}{|c|}{\cellcolor[HTML]{4A86E8}\textbf{State Category (27 OOV words)}} \\ \hline
\multicolumn{1}{|c|}{\cellcolor[HTML]{4A86E8}{\color[HTML]{091E42} Hard cost}} &
  {\color[HTML]{091E42} 33.3} &
  \multicolumn{1}{c|}{{\color[HTML]{091E42} 51.9}} &
  {\color[HTML]{091E42} 29.6} &
  \multicolumn{1}{c|}{{\color[HTML]{091E42} 55.6}} &
  {\color[HTML]{091E42} 37} &
  {\color[HTML]{091E42} 55.6} \\ \cline{1-1}
\multicolumn{1}{|c|}{\cellcolor[HTML]{4A86E8}{\color[HTML]{091E42} Acoustic}} &
  {\color[HTML]{091E42} 33.3} &
  \multicolumn{1}{c|}{{\color[HTML]{091E42} 51.9}} &
  {\color[HTML]{091E42} 29.6} &
  \multicolumn{1}{c|}{{\color[HTML]{091E42} 63}} &
  {\color[HTML]{091E42} 37} &
  {\color[HTML]{091E42} 55.6} \\ \cline{1-1}
\multicolumn{1}{|c|}{\cellcolor[HTML]{4A86E8}{\color[HTML]{091E42} Phonetic}} &
  {\color[HTML]{091E42} 33.3} &
  \multicolumn{1}{c|}{{\color[HTML]{091E42} 63}} &
  {\color[HTML]{091E42} \textbf{29.6}} &
  \multicolumn{1}{c|}{{\color[HTML]{091E42} \textbf{70.4}}} &
  {\color[HTML]{091E42} 37} &
  {\color[HTML]{091E42} 59.3} \\ \cline{1-1}
\multicolumn{1}{|c|}{\cellcolor[HTML]{4A86E8}{\color[HTML]{091E42} Append}} &
  {\color[HTML]{091E42} 33.3} &
  \multicolumn{1}{c|}{{\color[HTML]{091E42} 63}} &
  {\color[HTML]{091E42} \textbf{29.6}} &
  \multicolumn{1}{c|}{{\color[HTML]{091E42} \textbf{70.4}}} &
  {\color[HTML]{091E42} 37} &
  {\color[HTML]{091E42} 59.3} \\ \cline{1-1}
\multicolumn{1}{|c|}{\cellcolor[HTML]{4A86E8}{\color[HTML]{091E42} Weighted}} &
  {\color[HTML]{091E42} 33.3} &
  \multicolumn{1}{c|}{{\color[HTML]{091E42} 63}} &
  {\color[HTML]{091E42} \textbf{29.6}} &
  \multicolumn{1}{c|}{{\color[HTML]{091E42} \textbf{70.4}}} &
  {\color[HTML]{091E42} 37} &
  {\color[HTML]{091E42} 59.3} \\ \hline
\multicolumn{7}{|c|}{\cellcolor[HTML]{4A86E8}\textbf{City Category (370 OOV words)}} \\ \hline
\multicolumn{1}{|c|}{\cellcolor[HTML]{4A86E8}{\color[HTML]{091E42} Hard cost}} &
  {\color[HTML]{091E42} 6.5} &
  \multicolumn{1}{c|}{{\color[HTML]{091E42} 29.2}} &
  {\color[HTML]{091E42} 12.4} &
  \multicolumn{1}{c|}{{\color[HTML]{091E42} 30}} &
  {\color[HTML]{091E42} 11.6} &
  {\color[HTML]{091E42} 32.2} \\ \cline{1-1}
\multicolumn{1}{|c|}{\cellcolor[HTML]{4A86E8}{\color[HTML]{091E42} Acoustic}} &
  {\color[HTML]{091E42} 6.5} &
  \multicolumn{1}{c|}{{\color[HTML]{091E42} 55.4}} &
  {\color[HTML]{091E42} 12.4} &
  \multicolumn{1}{c|}{{\color[HTML]{091E42} 47}} &
  {\color[HTML]{091E42} 11.6} &
  {\color[HTML]{091E42} 52.2} \\ \cline{1-1}
\multicolumn{1}{|c|}{\cellcolor[HTML]{4A86E8}{\color[HTML]{091E42} Phonetic}} &
  {\color[HTML]{091E42} 6.5} &
  \multicolumn{1}{c|}{{\color[HTML]{091E42} 77.6}} &
  {\color[HTML]{091E42} 12.4} &
  \multicolumn{1}{c|}{{\color[HTML]{091E42} 74.1}} &
  {\color[HTML]{091E42} 11.6} &
  {\color[HTML]{091E42} 75.9} \\ \cline{1-1}
\multicolumn{1}{|c|}{\cellcolor[HTML]{4A86E8}{\color[HTML]{091E42} Append}} &
  {\color[HTML]{091E42} 6.5} &
  \multicolumn{1}{c|}{{\color[HTML]{091E42} 79.5}} &
  {\color[HTML]{091E42} 12.4} &
  \multicolumn{1}{c|}{{\color[HTML]{091E42} 75.7}} &
  {\color[HTML]{091E42} \textbf{11.6}} &
  {\color[HTML]{091E42} \textbf{77.6}} \\ \cline{1-1}
\multicolumn{1}{|c|}{\cellcolor[HTML]{4A86E8}{\color[HTML]{091E42} Weighted}} &
  {\color[HTML]{091E42} 6.5} &
  \multicolumn{1}{c|}{{\color[HTML]{091E42} 78.6}} &
  {\color[HTML]{091E42} 12.4} &
  \multicolumn{1}{c|}{{\color[HTML]{091E42} 75.9}} &
  {\color[HTML]{091E42} 11.6} &
  {\color[HTML]{091E42} 77} \\ \hline
\multicolumn{7}{|c|}{\cellcolor[HTML]{4A86E8}\textbf{Dealer Category (546 OOV words)}} \\ \hline
\multicolumn{1}{|c|}{\cellcolor[HTML]{4A86E8}{\color[HTML]{091E42} Hard cost}} &
  {\color[HTML]{091E42} 9} &
  \multicolumn{1}{c|}{{\color[HTML]{091E42} 83.2}} &
  {\color[HTML]{091E42} 0.7} &
  \multicolumn{1}{c|}{{\color[HTML]{091E42} 64.8}} &
  {\color[HTML]{091E42} 8.6} &
  {\color[HTML]{091E42} 83.3} \\ \cline{1-1}
\multicolumn{1}{|c|}{\cellcolor[HTML]{4A86E8}{\color[HTML]{091E42} Acoustic}} &
  {\color[HTML]{091E42} 9} &
  \multicolumn{1}{c|}{{\color[HTML]{091E42} 87}} &
  {\color[HTML]{091E42} 0.7} &
  \multicolumn{1}{c|}{{\color[HTML]{091E42} 90.1}} &
  {\color[HTML]{091E42} 8.6} &
  {\color[HTML]{091E42} 87.9} \\ \cline{1-1}
\multicolumn{1}{|c|}{\cellcolor[HTML]{4A86E8}{\color[HTML]{091E42} Phonetic}} &
  {\color[HTML]{091E42} 9} &
  \multicolumn{1}{c|}{{\color[HTML]{091E42} 89}} &
  {\color[HTML]{091E42} 0.7} &
  \multicolumn{1}{c|}{{\color[HTML]{091E42} 96.7}} &
  {\color[HTML]{091E42} 8.6} &
  {\color[HTML]{091E42} 89.4} \\ \cline{1-1}
\multicolumn{1}{|c|}{\cellcolor[HTML]{4A86E8}{\color[HTML]{091E42} Append}} &
  {\color[HTML]{091E42} \textbf{9}} &
  \multicolumn{1}{c|}{{\color[HTML]{091E42} \textbf{89.6}}} &
  {\color[HTML]{091E42} 0.7} &
  \multicolumn{1}{c|}{{\color[HTML]{091E42} 97.1}} &
  {\color[HTML]{091E42} 8.6} &
  {\color[HTML]{091E42} 89.7} \\ \cline{1-1}
\multicolumn{1}{|c|}{\cellcolor[HTML]{4A86E8}{\color[HTML]{091E42} Weighted}} &
  {\color[HTML]{091E42} \textbf{9}} &
  \multicolumn{1}{c|}{{\color[HTML]{091E42} \textbf{89.6}}} &
  {\color[HTML]{091E42} 0.7} &
  \multicolumn{1}{c|}{{\color[HTML]{091E42} 97.1}} &
  {\color[HTML]{091E42} 8.6} &
  {\color[HTML]{091E42} 89.7} \\ \hline
\end{tabular}
\end{table}

The results of the sentence level data sets are shown in Table~\ref{tab:sentence_results}. We used a window length of $\pm500$ ms from the start and end of the reference context word for searching context words within a decoded hypothesis. We can see from the table that more than 50\% of the OOVs are recovered correctly. We also experimented with different acoustic scales and the cost threshold for OOV recovery. We found that increasing the acoustic scale makes a recovery better, maxing out at a scale of 10. However, many unwanted additional words are decoded and create more confusion, making it harder to recover OOV context words. As we increase the cost threshold for OOV recovery beyond some value, many words from the OOV list are getting updated by Pass $2$, which will create a problem in combining the text from both ASR decodes into the final transcription. Due to modeling errors, neither the ground truth nor the decoded sentence would have the perfect time alignment. Further, the margin of error can be avoided by optimizing this searching window. To fix the hyper-parameter, we varied the searching window lengths from 100 to 500 ms and found that the 500 ms window length gave the best result.

% \usepackage{graphicx}
\begin{table}[]
\caption{Experiment 3: OOV recovery rate results of multiple cost functions for sentence level data set (D2) with multiple OOV context word list}
\label{tab:sentence_results}
\centering
\resizebox{\textwidth}{!}{%
\begin{tabular}{|c|c|c|c|c|c|c|}
\hline
\rowcolor[HTML]{FBBC04} 
\textbf{\begin{tabular}[c]{@{}c@{}}Cost Function \\ Method\end{tabular}} &
  \textbf{\begin{tabular}[c]{@{}c@{}}Window \\ length (ms)\end{tabular}} &
  \textbf{\begin{tabular}[c]{@{}c@{}}Cost \\ Threshold\end{tabular}} &
  \textbf{\begin{tabular}[c]{@{}c@{}}Acoustic \\ Scale\end{tabular}} &
  \textbf{\begin{tabular}[c]{@{}c@{}}Graph \\ scale\end{tabular}} &
  \textbf{Pass 1 (\%)} &
  \textbf{Pass 2 (\%)} \\ \hline
Hard cost         & 500          & 10          & 5           & 50          & 18          & 22          \\ \hline
Acoustic          & 500          & 10          & 5           & 50          & 26          & 22          \\ \hline
Phonetic          & 500          & 10          & 5           & 50          & 26          & 28          \\ \hline
Append            & 500          & 10          & 5           & 50          & 26          & 28          \\ \hline
\textbf{Weighted} & \textbf{500} & \textbf{10} & \textbf{5}  & \textbf{50} & \textbf{26} & \textbf{30} \\ \hline
Weighted          & 500          & 5           & 5           & 50          & 26          & 24          \\ \hline
Weighted          & 200          & 5           & 5           & 50          & 22          & 22          \\ \hline
Weighted          & 500          & 3           & 5           & 50          & 26          & 16          \\ \hline
% Weighted          & 400          & 3           & 5           & 50          & 24          & 14          \\ \hline
Weighted          & 300          & 3           & 5           & 50          & 22          & 14          \\ \hline
% Weighted          & 200          & 3           & 5           & 50          & 22          & 12          \\ \hline
Weighted          & 100          & 3           & 5           & 50          & 24          & 8           \\ \hline
\textbf{Weighted} & \textbf{500} & \textbf{10} & \textbf{10} & \textbf{50} & \textbf{24} & \textbf{34} \\ \hline
% Weighted          & 500          & 10          & 5           & 100         & 24          & 30          \\ \hline
% Weighted          & 500          & 10          & 10          & 100         & 24          & 34          \\ \hline
% Weighted          & 500          & 10          & 25          & 25          & 24          & 32          \\ \hline
\end{tabular}%
}
\end{table}

\section{Conclusion}
This paper proposed a post-processing technique to recover context-based OOV words. The proposed method is a simple approach to recovering important words, given that the list of context words is known. We focused mainly on finding a good phone cost function to recover OOVs. We explored acoustic and phonetic similarity-based cost functions. The experimental evaluation shows that the proposed method gives OOV context word recovery of more than $50\%$ across multiple categories. We have only explored phone substitution manipulation keeping the insertion and deletion costs fixed. The insertion and deletion costs could be adjusted for future work. For example, consonant deletions can be penalized less compared to vowel deletions. 

\bibliographystyle{splncs04}
\bibliography{mybib}
\end{document}